\documentclass[seceq]{ptptex}
\usepackage{amsbsy,amssymb,latexsym,amsmath,epsfig}

\def\CF{{\cal F}}

\def\CQ{{\mathcal Q}}

\def\CL{{\cal L}}

\def\CV{{\cal V}}

\def\1{\mathbb I}
\def\im{{\rm Im}\,}
\def\re{{\rm Re}\,}

\def\CL{{\mathcal L}}
\def\CM{{\mathcal M}}

\def\CD{{\mathcal D}}

\def\CE{{\mathcal E}}
\def\CM{{\mathcal M}}

\def\CF{{\mathcal F}}

\def\Tr{{\rm Tr}}
\def\tr{{\rm tr}}

\def\CP{{\cal P}}

\newcommand{\bbR}{{\mathbb R}}

\newcommand\CW{{\mathcal W}}

\newcommand\CN{{\mathcal N}}
\newcommand\fR{{\mathfrak R}}
\renewcommand{\overleftrightarrow}[1]{\stackrel{\leftrightarrow}{#1}}

\newcommand\bR{{\mathbb R}}

\newcommand\rangled{\rangle\! \rangle}
\newcommand\langled{\langle\!\langle}

\title{
\vspace{-22.8mm}

\textmd{{\small 
\hfill 
OCU-PHYS 242~~~~~
\\\hfill 
OIQP-05-17~~~~~
}}\\
Supersymmetric $U(N)$ Gauge Model
and\\
Partial Breaking of $\CN=2$ Supersymmetry
\footnote{Talk given by H.I.
at 
the international workshop 
``Frontier of Quantum Physics'', 
February 17-19, 2005, Yukawa Institute for Theoretical Physics,
Kyoto University, Japan,
and talk given by M.S.
at 
the workshop 
``Progress of Quantum Field Theory and String Theory'', 
February 6-7, 2006, Osaka City University Advanced Mathematical Institute (OCAMI), 
Japan.}
}

\author{
K. \textsc{Fujiwara}$^1$~,~~
H. \textsc{Itoyama}$^1$~~\&~~
M. \textsc{Sakaguchi}$^2$
}

\inst{
$^1$~Department of Mathematics and Physics,
Graduate School of Science,\\
Osaka City University\\
3-3-138, Sugimoto, Sumiyoshi-ku, Osaka, 558-8585, Japan
\medskip

$^2$~Okayama Institute for Quantum Physics,
\\~~~~
1-9-1 Kyoyama, Okayama 700-0015, Japan
}

\abst{
We review the construction of the $\CN=2$ $U(N)$ gauge model
and the analysis of vacua of the model.
On the vacua,
$\mathcal{N}=2$ supersymmetry is spontaneously
broken to $\mathcal{N}=1$, 
and the gauge symmetry is broken to a product gauge group
$\displaystyle{\prod _{i=1}^n U(N_i)}$. 
The masses of the supermultiplets appearing on the 
$\mathcal{N}=1$ vacua
are given.
We provide a manifestly $\CN=2$ supersymmetric formulation
of the $U(N)$ gauge model coupled with $\CN=2$ hypermultiplets,
and show that $\CN=2$ supersymmetry is partially broken down to $\CN=1$ spontaneously.
}

\begin{document}

\maketitle

\section{Introduction}

Until mid nineties,
partial breaking of global extended supersymmetries
was thought not to be possible.
The statement is as follows:
\begin{quote}
\textit{
Start with the $\CN$-extended supersymmetry algebra
\begin{eqnarray}
{\left\{ \bar{Q}_{\alpha}^I,Q_{J\dot{\alpha}} \right\}
=2\delta_{\alpha \dot{\alpha}}\delta_J^{I} H~,~~~~
I,J=1,...,\CN~.
\nonumber
}\end{eqnarray}
This implies that
\begin{eqnarray}
2H=\sum_{\dot\alpha}
||Q_{I\dot \alpha}|0\rangle||^2
~~~~~\forall I~.
\nonumber
\end{eqnarray}
If $ Q_{I}|0 \rangle=0$ for some $I$,
then  $H=0$.
This implies that $Q_{I}|0 \rangle=0$ for all $I$
because the right hand side is positive definite.
On the other hand, 
if $ Q_{I}|0 \rangle\neq 0$ for some $I$,
then  $H>0$.
This implies that $Q_{I}|0 \rangle\neq 0$ for all $I$.
Thus, in an $\CN$-extended global supersymmetric theory,
either all or no supersymmetry is spontaneously broken.
}
\end{quote}
Obviously, this does not apply to
 \textit{local} supersymmetry,
 because
the Hilbert space metric is not
positive definite.
For the \textit{rigid} case
a loophole for this statement 
is to use the supercurrent algebra,
and the most general form is 
\begin{eqnarray}
{\left\{ \bar{Q}_{\dot{\alpha}}^J,\mathcal{S}_{\alpha I}^m (x) \right\}=
2(\sigma^n)_{\alpha \dot{\alpha}} {\delta_I^J} T_n^m(x)
+(\sigma^m)_{\alpha \dot{\alpha}} }{C_I{}^J}
\label{supercurrent algebra}
\end{eqnarray}
where $\mathcal{S}_{\alpha I}^m$ are extended supercurrents,
$T_n^m$ is the stress-energy tensor 
and {$C_I{}^J$} is a field independent constant matrix,
which is permitted by the constraint
for the Jacobi identities~\cite{Lopuszanski:1978df}.
The new term does not modify the supersymmetry algebra on the fields.
Partial supersymmetry breaking discussed in the
present paper corresponds to this case.

Besides active researches on
the \textit{non-linear} realization of extended supersymmetry
in the partially broken phase,
a model
in which \textit{linearly} realized $\CN=2$ supersymmetry is
partially broken to $\CN=1$ spontaneously
was given by Antoniadis, Partouche and Taylor (APT)
\cite{APT} (see also \citen{central_charge}\citen{PP}\citen{Ivanov:1997mt}).
APT model is
$\mathcal{N}=2$ supersymmetric, self-interacting {$U(1)$} model
with one (or several) abelian $\mathcal{N}=2$ vector multiplet(s) 
$\mathcal A$~\cite{N=2_constraint}
with electric \& magnetic Fayet-Iliopoulos (FI) terms.
In \citen{FIS1}\citen{FIS2},
we have generalized this model to the {$U(N)$} gauge model
and shown that the $\CN=2$ supersymmetry is partially broken to
$\CN=1$ spontaneously.
Further in \citen{FIS3},
we have analyzed the vacua with broken gauge symmetry
and revealed the $\CN=1$ supermultiplets on the vacua.
In addition,
a manifestly $\CN=2$ supersymmetric formulation of the
$U(N)$ gauge model coupled with/without $\CN=2$ hypermultiplets
was given in {\citen{FIS4}}
by using unconstrained $\CN=2$ superfields
on harmonic superspace\cite{HS}.
We introduce the magnetic Fayet-Iliopoulos term
so as to shift the auxiliary field in $\CN=2$ vector multiplet
by an imaginary constant.
We find that
in presence of hypermultiplets in the fundamental representation
of $U(N)$,
the magnetic FI term 
develops an additional term
which overcomes the difficulty
\cite{PP}\cite{Ivanov:1997mt}
 in coupling
fundamental hypermultiplets with the APT model.
In these models,
the renormalizability is not imposed and the
prepotential $\mathcal{F}$ appears from the beginning.
Thus, our model should be regarded as a low-energy effective action
for systems given by $\CN=2$ bare actions spontaneously broken to $\CN=1$.
See \citen{SW}\citen{string}\citen{DV}\citen{Cachazo:2002ry}
for related references.

This paper is organized as follows.
In the next section,  $\CN=2$ $U(N)$ gauge model is constructed
by requiring $\CN=1$ $U(N)$ gauge model to be $\fR$-invariant.
The $\fR$-action is composed of the discrete element of
the $SU(2)$ $R$-symmetry,
the automorphism of $\CN=2$,
and a sign flip of the FI 
D-term.
$\CN=2$ supersymmetry transformations are given in section 3.
In section 4, we analyze the vacua of the model,
and find that $\CN=2$ supersymmetry
and the $U(N)$ gauge symmetry
are partially broken to $\CN=1$ and $\prod_iU(N_i)$, respectively.
We clarify the mass spectrum of the model,
and reveal the $\CN=1$ supermultiplets on the vacua
in section 5.
In section 6, we discuss the $\CN=2$ supercurrents
and the ``central charge'' in (\ref{supercurrent algebra}).
The last section is devoted to a manifestly $\CN=2$ supersymmetric formulation
of the $U(N)$ gauge model coupled with $\CN=2$ hypermultiplets,
and we show that $\CN=2$ supersymmetry is spontaneously broken to $\CN=1$.

\section{$\CN=2$ $U(N)$ gauge model}
We introduce
an $\CN=1$ chiral superfield,
$\Phi(x^m,\theta) =  \sum_{a=0}^{N^2-1} \Phi ^a t_a$,
where $N\times N$ hermitian matrices
$t_a$ ($a=0,...,N^2-1$)
generate $u(N)$ algebra,
$[t_a, t_b]=if^c_{ab}t_c,~
\tr (t_at_b)=\frac{1}{2}\delta_{ab}$,
 ($t_0$ generates the overall u(1)).
The kinetic term for $A$ ($\Phi\ni (A,\psi,F)$) we use is given by
the K\"ahler potential for the special K\"ahler geometry
\begin{eqnarray}
\CL_{K}=
\int d^2\theta d^2\bar\theta ~K(\Phi^a,\Phi^{*a}),~~~~
 K=\frac{i}{2} (\Phi^a \CF^*_a
-\Phi^{{*a}} \CF_a),
\end{eqnarray}
where $\CF_a=\frac{\partial\CF}{\partial \Phi^a}$
and $\CF$ is an analytic function of $\Phi$.
The K\"ahler metric {$g_{ab^*}=\partial_a\partial_{b^*}K|=\im \CF_{ab}|$} admits $U(N)$ isometry generated by holomorphic Killing vectors~
$k_a=k_a{}^b\partial_b$ and
$k^*_a=k^*_a{}^b\partial_b$ with
\begin{eqnarray}
k_a{}^b=-ig^{bc^*}\partial_{c^*}\CP_a,~~~
k^*_a{}^b=+ig^{cb^*}\partial_{c}\CP_a,
\end{eqnarray}
where $\CP_a$ is called as the Killing potential.
In the present case, $\CP_a$ is give by
\begin{eqnarray}
\CP_a= -\frac{1}{2}(\CF_bf^b_{ac}A^{*c}+\CF^*_bf^b_{ac}A^c)~.
\end{eqnarray}
$A^a$ and  $\CF_b$ transform in the adjoint representation of $U(N)$
\begin{eqnarray}
k_a^c\partial_cA^b
=f^b_{ac}A^c,~~~~
k_a^c\partial_c\CF_b=-f^c_{ab}\CF_c~.
\end{eqnarray}
To gauge this isometry, we introduce
an
$\CN=1$ vector superfields,
$
 V(x^m,\theta,\bar{\theta})= \sum_{a=0}^{N^2-1} V^a t_a
$,~
$V^a\ni (v_m^a,\lambda^a,D^a)$.
The $U(N)$ gauging is accomplished by adding
\cite{Bagger:1982fn}\cite{Hull:1985pq}
\begin{eqnarray}
\CL_\Gamma=\int d^2\theta d^2\bar\theta ~ \Gamma,~~~~~~
\Gamma=
\left[
\int^1_0d\alpha ~e^{\frac{i}{2}\alpha  v^a(k_a-k_a^*)}v^c\CP_c
\right]_{v^a\to V^a}~.
\end{eqnarray}
For the gauge kinetic term,
we introduce
\begin{eqnarray}
\CL_{\CW^2}=
-\frac{i}{4}\int d^2\theta ~ \tau_{ab}\,\CW^a\CW^b + c.c.~~~~~
{\mathcal W}_{\alpha}=-\frac{1}{4} \bar{D} \bar{D} e^{-V} D_{\alpha}
e^V={\mathcal W}_{\alpha}^a t_a
\end{eqnarray}
where $\tau_{ab}(\Phi)$ is an analytic function of $\Phi$.
In addition, we introduce
a gauge invariant 
superpotential term and the FI D-term \cite{FI}
\begin{eqnarray}
\CL_W=
 \int d^2\theta~ W(\Phi) +c.c.~,~~~~~
\CL_{D}=\xi \int d^2 \theta d^2 \bar{\theta} V^0 =
\sqrt{2}\xi D^0.
\end{eqnarray}
In summary, 
the total Lagrangian of 
the $\CN=1$ $U(N)$
gauge model is
\begin{eqnarray}
{\CL=\CL_K+\CL_\Gamma+\CL_{\CW^2}+\CL_W+\CL_D}~.
\end{eqnarray}

The auxiliary fields are evaluated as
\begin{eqnarray}
  \begin{array}{ll}
D^{a}=
{\hat{D}^a}
-(\tau_2^{-1})^{ab}\left(
\frac{1}{2}\CP_b + \sqrt{2}\xi\delta_{b}^0
\right),       
&{
\hat{D}^a=-\frac{\sqrt{2}}{4}(\tau_2^{-1})^{ab} \left( \partial_d\tau_{bc}\psi^d\lambda^c+\partial_{d^*}\tau^*_{bc}\bar\psi^d\bar\lambda^c \right)},    \\
F^a={\hat{F}^a}
-g^{ab^*}\partial_{b^*} W^*,       
&{
\hat{F}^a=-g^{ab^*} \left( -\frac{i}{4}\partial_{b^*}\tau^*_{cd}\bar\lambda^c\bar\lambda^d-\frac{1}{2}g_{cb^*,d}\psi^c\psi^d \right)},
    \\
F^{*a}={\hat{F}^{*a}}
-g^{ba^*}\partial_{b} W~,       
&{\hat{F}^{*a}=-g^{ba^*} \left( \frac{i}{4}\partial_{b}\tau_{cd}\lambda^c\lambda^d-\frac{1}{2}g_{bc^*,d^*}\bar\psi^c\bar\psi^d \right)}~,    \\
  \end{array}
\nonumber\\&&
\end{eqnarray}
where $(\tau_2)_{ab}=\im \tau_{ab}$, and $\hat{D}^a$, $\hat{F}^a$ and $\hat{F}^{*a}$ 
are fermion bilinear terms.
Eliminating auxiliary fields by using
the above expressions
and defining covariant derivatives by
\begin{eqnarray}
&&\CD_m\Psi^a=\partial_m\Psi^a-\frac{1}{2}f_{bc}^{a}v_m^b\Psi^c~,~~~
\Psi=\{A, \psi, \lambda\},
\nonumber\\
&&\CD_m'\psi^a=
\CD_m\psi^a+\Gamma^a_{bc}\CD_mA^b\psi^c~,~~~~~
v_{mn}^a=
\partial_mv_n^a
-\partial_nv_m^a
-\frac{1}{2}f^a_{bc}v^b_mv^c_n~,
\end{eqnarray}
the total action is summarized as 
$
\CL
=
\CL_{\rm{kin}}
+\CL_{\rm{pot}}
+\CL_{\rm{Pauli}}
+\CL_{\rm{Yukawa}}
+\CL_{\rm{fermi^4}} $
with
\begin{eqnarray}
\CL_{\rm{kin}}&=&
-g_{ab^*}\mathcal{D}_m A^a \mathcal{D}^m A^{*b}
-\frac{1}{4} (\tau_2)_{ab} v_{mn}^a v^{bmn}
-\frac{1}{8} \re \tau_{ab}\, \epsilon^{mnpq}v_{mn}^a v_{pq}^b
\nonumber\\&&
+[
-\frac{1}{2} \tau_{ab} \lambda^a \sigma^m \mathcal{D}_m
\bar{\lambda}^b
-\frac{i}{2}g_{ab^*} \psi^a\sigma^m \mathcal{D}'_m \bar{\psi}^b
~~+~~c.c.
]~,
\\
\CL_{\rm{pot}}&=&
-\frac{1}{2} \left( \tau_2^{-1} \right)^{ab} \left(
\frac{1}{2}\CP_{a} + \sqrt{2} \xi\delta_{a}^0
\right)
\left(
\frac{1}{2}\CP_{b} + \sqrt{2} \xi\delta_{b}^0
\right)
-g^{ab^*}\partial_a W \partial_{b^*} {W}^*,
\\
\CL_{\rm{Pauli}}&=&
[
-i\frac{\sqrt{2}}{8} \partial_c \tau_{ab} \psi^c \sigma^n
\bar{\sigma}^m \lambda^a v_{mn}^b
~~+~~c.c.
]~,
\\
\CL_{\rm{Yukawa}}&=&
[
-\frac{1}{2}\left(
 \partial_a\partial_b W
 -g^{cd^*}\partial_cW g_{ad^*,b}
 \right) \psi^a \psi^b
-\frac{i}{4}g^{cd^*}\partial_{d^*}W^*\partial_c\tau_{ab}\lambda^a\lambda^b
~~+~~c.c.
]~
\nonumber \\&&
+[\left(
\frac{1}{\sqrt{2}} g_{ac^*} 
k_b^*{}^{c}
-\frac{\sqrt{2}}{4} \left( \tau_2^{-1} \right)^{cd}
\left(\frac{1}{2}\CP_{d} + \sqrt{2} \xi\delta_{d}^0 \right)
\partial_a \tau_{bc}
\right)\psi^a \lambda^b
~~+~~c.c.]
~,
\nonumber \\&&
\\
\CL_{\rm{fermi^4}}&=&
[
-\frac{i}{8}\partial_c \partial_d \tau_{ab}
\psi^c \psi^d \lambda^a\lambda^b
~~+~~c.c.]
~
\\&&
-\frac{1}{16}\left(\tau_2^{-1} \right)^{ab}
\left( \partial_d \tau_{ac} \psi^d \lambda^c
+\partial_{d^*} {\tau}^*_{ac} \bar{\psi}^d \bar{\lambda}^c
\right)
\left( \partial_f \tau_{be} \psi^f \lambda^e
+\partial_{f^*} {\tau}^*_{be} \bar{\psi}^f \bar{\lambda}^e
\right)
\nonumber \\&&
-g^{ab^*}
\left(
\frac{i}{4}\partial_a \tau_{cd} \lambda^c\lambda^d
-\frac{1}{2}g_{ac^*,d^*}\bar{\psi}^c \bar{\psi}^d
\right)
\left(
-\frac{i}{4}\partial_{b^*}{\tau}^*_{ef}
\bar{\lambda}^e\bar{\lambda}^f
-\frac{1}{2}g_{eb^*,f} \psi^e \psi^f
\right).
\nonumber
\end{eqnarray}

Now, we require that the action
is invariant under $\fR$-action,
{$\fR: S \to S$}. 
The $\fR$-action is composed of a discrete element of the $SU(2)$ R-symmetry,
automorphism of $\CN=2$,
and a sign flip of the FI parameter
\begin{eqnarray}
R:\left(
  \begin{array}{c}
  \lambda^a     \\
  \psi^a     \\
  \end{array}
\right)
\longrightarrow
\left(
  \begin{array}{c}
  \psi^a     \\
  -\lambda^a     \\
  \end{array}
\right)
~~~~~~\&~~~~~~~
R_\xi:
\xi \to -\xi~,
\end{eqnarray}
so that $S^{(+\xi)}\xrightarrow{R} S^{(-\xi)}\xrightarrow{R_\xi}S^{(+\xi)}$
where we have made the sign of the FI parameter manifest.
This ensures the $\CN=2$ supersymmetry of our action as follows
(see also Appendix A in \citen{FIS1}).
By construction, the action is invariant under the first supersymmetry 
$\delta_{\eta_1}S^{(+\xi)}=0$.
Acting $\fR$ on it, we have
\begin{eqnarray}
\delta_{\eta_1}S^{(+\xi)}=0
~~\xrightarrow{R}~~
R(\delta_{\eta_1})S^{(-\xi)}=0
~~\xrightarrow{R_\xi}~~
\fR(\delta_{\eta_1})S^{(+\xi)}=0~,
\end{eqnarray}
which implies that
the resulting $\fR$-invariant action is invariant under the second supersymmetry
$\delta_{\eta_2}\equiv\fR (\delta_{\eta_1})$ as well.

In \citen{FIS1}, we find that the action
is invariant under $\fR$-action,
and thus $\CN=2$ supersymmetric, if
\begin{eqnarray}
\tau_{ab}=\CF_{ab}~,~~~
W=eA^0+m\CF_0~.
\end{eqnarray}
The $\fR$-action on the auxiliary fields are
\begin{eqnarray}
F^a+g^{ac^*}\partial_{c^*} W^*
~\to~
F^{*b}+g^{db^*}\partial_{d}W~,~~
D^c+\frac{1}{2}g^{cd}\CP_d
~\to~
-( D^c+\frac{1}{2}g^{cd}\CP_d)
~~~~~
\end{eqnarray}
or equivalently, 
$
\hat F^a
\to
\hat F^{*a}~,~
\hat D^a
\to
- \hat D^a
$,
which are consistent with the $\fR$-action on the fermions.

\section{$\CN=2$ supersymmetry transformation}

We construct the second supersymmetry transformation
by applying the $\fR$-action on the first supersymmetry transformation
\begin{eqnarray}
&&
\delta_{\eta_1} A^a=
\sqrt{2}\eta_1\psi^a~,~~~
\delta_{\eta_1} \psi^a=
 i\sqrt{2}\sigma^m\bar\eta_1\CD_mA^a
 +\sqrt{2}\eta_1 F^a~,
\nonumber\\&&
\delta_{\eta_1} v_m^a=
 i\eta_1\sigma_m\bar\lambda^a-i\lambda^a\sigma_m\bar\eta_1~,~~~~
 \delta_{\eta_1} \lambda^a=
\sigma^{mn}
\eta_1 v_{mn}^a+i\eta_1 D^a~.
\end{eqnarray}
The $\fR$-action on the fields is summarized as
\begin{eqnarray}
&&
\boldsymbol{\lambda}_I^{\ a} \equiv 
\left(\!\!\!
\begin{array}{c}
\lambda^a \\
\psi^a
\end{array}
\!\!\!\right)
\longrightarrow 
\left(\!\!\!
\begin{array}{c}
\psi^a \\
-\lambda^a
\end{array}
\!\!\!\right)
 \equiv
\boldsymbol{\lambda}^{Ia}=\epsilon^{IJ}\boldsymbol{\lambda}_J{}^a
~,~~~~~~\epsilon^{12}=\epsilon_{21}=1~,
\\&&F^a=\hat F^a-g^{ab^*}\partial_{b^*}W^*
\longrightarrow \hat F^{*a}-g^{ab^*}\partial_{b^*}W^*~,
\\&&
D^a=\hat D^a-(\tau_2^{-1})^{ab}(\frac{1}{2}\CP_b+\sqrt{2}\xi\delta_b^0)
\longrightarrow  -\hat D^a-(\tau_2^{-1})^{ab}(\frac{1}{2}\CP_b-\sqrt{2}\xi\delta_b^0)
\end{eqnarray}
while $A^a$ and $v_m^a$ are $\fR$-invariant.
For $\boldsymbol{\delta} A^a$
and $\boldsymbol{\delta}v_m^a$ 
($\boldsymbol{\delta}=\delta_{\eta_1}+\delta_{\eta_2}$) to be
$\fR$-invariant,
the supersymmetry parameters $\eta_1$ and $\eta_2$ must form
a doublet
$
\boldsymbol{\eta}_I \equiv 
\left(\!\!\!
\begin{array}{c}
\eta_1 \\
\eta_2
\end{array}\!\!\!
\right)
\to
\left(\!\!\!
\begin{array}{c}
\eta_2 \\
-\eta_1
\end{array}\!\!\!
\right)
 \equiv
\boldsymbol{\eta}^{I}=\epsilon^{IJ}\boldsymbol{\eta}_J~.
$
As a result, 
we obtain the $\CN=2$ supersymmetry transformation
\begin{eqnarray}
\boldsymbol{\delta} A^a &=&
\sqrt{2} \boldsymbol{\eta}_J \boldsymbol{\lambda}^{Ja}~,\\
\boldsymbol{\delta} v_m^{a} &=&
i \boldsymbol{\eta}_J \sigma_m \bar{\boldsymbol{\lambda}}^{Ja}   
-i \boldsymbol{\lambda}_J^{\ a}{\sigma}_m\bar{\boldsymbol{\eta}}^J ~,\\
\boldsymbol{\delta \lambda}_J^{\ a} &=& 
(\sigma^{mn} \boldsymbol{\eta}_J)v_{mn}^{a}
+\sqrt{2}i(\sigma^m \bar{\boldsymbol{\eta}}_J) \mathcal{D}_m A^a
+i(\boldsymbol{\tau} \cdot \boldsymbol{D}^a)_J{}^K \boldsymbol{\eta}_K
-\frac{1}{2} \boldsymbol{\eta}_J f^a_{\ bc} A^{*b} A^c,~~~
\end{eqnarray}
where $\boldsymbol{\tau}$ are Pauli matrices, and 3-dimensional vector
$\boldsymbol{D}^a$ is given by
\begin{eqnarray}
\boldsymbol{D}^a &=&\hat {\boldsymbol{D}}^a -\sqrt{2} g^{ab^*}
\partial_{b^*}
\left( \boldsymbol{\mathcal{E}}A^{*0}+\boldsymbol{\mathcal{M}}
{\mathcal{F}}_0^* \right),
\\
\hat{\boldsymbol{D}}^a&=&
(\hat{D}_1^a, \ \hat{D}_2^a, \ \hat{D}_3^a)
=(\sqrt{2}\im \hat F^a, -\sqrt{2}\re \hat F^a, \hat{D}^a ), 
\\
\boldsymbol{\mathcal{E}}&=&(0,\ -e,\ \xi)~,~~~
\boldsymbol{\mathcal{M}}=(0,\ -m,\ 0)~.
\end{eqnarray}
This would be $SU(2)$ covariant if 
$\boldsymbol{D}^a$
transformed as a triplet. 
In reality, the rigid $SU(2)$ has been gauge fixed
by making  $\boldsymbol{\mathcal{E}}$ and $\boldsymbol{\mathcal{M}}$ point 
to specific directions. 

Under the symplectic transformation,
$\Omega=\dbinom{A^0}{\CF_0}\to \Lambda\Omega$,
$\Lambda\in Sp(2,\bbR)$,
$\dbinom{-\CM}{\CE}$ changes to
$\dbinom{-\CM'}{\CE'}=\Lambda^{-1}\dbinom{-\CM}{\CE}$.
So the electric and magnetic charges are interchanged $(\CE',\CM')=(\CM,-\CE)$
when $\Lambda=
\left(
  \begin{array}{@{\,}cc@{\,}}
    0   &-1    \\
      1 &0    \\
  \end{array}
\right)
$.
This explains the name of the electric and magnetic FI terms
(see \S7).

The $\boldsymbol{D}^a$ is
not real but complex,
$
\textrm{Im } \boldsymbol{D}^a = \delta^a_{\ 0} (-\sqrt{2} m)
\left(
0,
1,
0
\right). 
$
As is seen in subsection 4.2,
this is necessary for the partial supersymmetry breaking.


\section{Analysis of vacua}
We examine vacua of the model 
and find that the $\CN=2$
supersymmetry and the $U(N)$ gauge symmetry are partially broken
to $\CN=1$ and $\prod_iU(N_i)$, respectively.

The scalar potential of our model is given by
$\CV\equiv-\CL_{\rm pot}$
\begin{eqnarray}
\displaystyle \CV=
\frac{1}{8}g_{ab}
\CP^a\CP^b
+\frac{1}{2}g_{ab}
\boldsymbol{\widetilde D}^a\cdot \boldsymbol{\widetilde D}^{*b}
~,
~~
\left\{
  \begin{array}{l}
\CP^a\equiv g^{ab}\CP_b=-if^a_{bc}A^{*b}A^c~,       \\
\boldsymbol{\widetilde D}^a\equiv-\sqrt{2}g^{ab^*}
\partial_{b^*}(\boldsymbol{\CE}A^{*0}+\boldsymbol{\CM}\CF^*_0)
\\
~~~~\,
=
\sqrt{2}g^{ab^*}(0,\partial_{b^*}W^*,-\xi\delta_b^0)    ~.   \\
  \end{array}
\right.
\end{eqnarray}
We require the positivity of the metric.
The first term vanish when $\langle A^r\rangle=0$,
where $A=A^at_a\equiv A^it_i+A^rt_r$
with {$t_i(t_r)\in$ (non-)Cartan}.
The vacua are specified by\cite{FIS2}
\begin{eqnarray}
\langle\frac{\partial\CV}{\partial A^a}\rangle
=
\frac{i}{4}\langle \CF_{abc}\boldsymbol{D}^b\cdot\boldsymbol{D}^c\rangle
=0~.
\end{eqnarray}
This determines the vacuum expectation value $\langled A^{i}\rangled$.

Let $E_{\underline{i}\underline{j}}$ ($\underline{i},\underline{j}=1,\cdots,N$) 
be the fundamental matrix,
which has $1$ at the $(\underline{i},\underline{j})$-component
and $0$ otherwise.
u(N) generators are given by
\begin{eqnarray}
\mbox{Cartan}&:&H_{\underline{i}}=E_{\underline{i}\underline{i}}~,
~~~~~~~\hfill ~\tr(H_{\underline{i}})^2=1\\
\mbox{non-Cartan}&:&
\left\{
  \begin{array}{l}
E_{\underline{i}\underline{j}}^+
=
\frac{1}{2}(E_{\underline{i}\underline{j}}+E_{\underline{j}\underline{i}})       \\
E_{\underline{i}\underline{j}}^-=
-\frac{i}{2}(E_{\underline{i}\underline{j}}-E_{\underline{j}\underline{i}})       \\
  \end{array}
\right.~(\underline{i}\neq \underline{j})~,~~
E_{\underline{i}\underline{j}}^\pm=\pm E_{\underline{i}\underline{j}}^\pm~,~
\tr(E_{\underline{i}\underline{j}}^\pm)^2=\frac{1}{2}~
~~~~
\end{eqnarray}
and $A$ is expanded as
{$ 
A=A^it_i+A^rt_r~
=A^{\underline{i}}H_{\underline{i}}
+\frac{1}{2}(A^{\underline{i}\underline{j}}_+E^+_{\underline{i}\underline{j}}
+A^{\underline{i}\underline{j}}_-E^-_{\underline{i}\underline{j}})$
}
with $A^{\underline{i}\underline{j}}_\pm=\pm A^{\underline{i}\underline{j}}_\pm$~.
The ordinary Cartan generators $t_i$ and $H_{\underline{i}}$ above are related by
{{$t_i=O_i{}^{\underline{j}}H_{\underline{j}}$}}.

For concreteness, we consider the prepotential
\begin{eqnarray}
\CF=\sum_{\ell=0}\frac{C_\ell}{\ell !}\Tr\Phi^\ell
~.
\end{eqnarray}
Non-vanishing $\langle \CF_{ab}\rangle$ and $\langle\CF_{abc}\rangle$  are
\begin{eqnarray}
\langle\CF_{\underline{i}\underline{i}}\rangle~,~~
\langle\CF_{\pm\underline{i}\underline{j},\pm\underline{i}\underline{j}}\rangle
\equiv\langle{\partial^2\CF}/
{\partial\Phi^{\underline{i}\underline{j}}_\pm
\partial\Phi^{\underline{i}\underline{j}}_\pm}\rangle~,~~
\langle\CF_{\underline{i}\underline{i}\underline{i}}\rangle~,~~
\langle\CF_{\underline{i},\pm\underline{i}\underline{j},\pm\underline{i}\underline{j}}\rangle
~,~~
\langle\CF_{\underline{j},\pm\underline{i}\underline{j},\pm\underline{i}\underline{j}}\rangle
~,~~~
\end{eqnarray}
and so the metric {\it $\langle g_{ab}\rangle$ is diagonal}.
The vacuum condition reduces to
\begin{eqnarray}
0=\langle \CF_{\underline{i}\underline{i}\underline{i}}\boldsymbol{D}^{\underline{i}}
\cdot
\boldsymbol{D}^{\underline{i}}\rangle~~~~
\forall  \underline{i}
\end{eqnarray}
because $\langle \boldsymbol{D}^r\rangle\sim
\langle
g^{rs}(\boldsymbol{\CE}\delta^0_s+\CM\CF_{0s}^*)
\rangle=0~.$
The points specified by $\langle \CF_{\underline{i}\underline{i}\underline{i}}\rangle=0$
are not stable vacua because 
$\langle\partial_{\underline{i}}\partial_{\underline{i}^*}\CV\rangle=0$.
At the stable vacua, we have
\begin{eqnarray}
{
\langle \boldsymbol{D}_{\underline{i}}
\cdot
\boldsymbol{D}_{\underline{i}}\rangle=0
~~~
\mbox{where}~~~
\langle\boldsymbol{D}_{\underline{i}}\rangle
=O_{\underline{i}}{}^j\langle\boldsymbol{D}_{j}\rangle
=\frac{2}{\sqrt{N}}\left(
0,\,
e+\frac{1}{2}m\langle\CF_{\underline{i}\underline{i}}^*\rangle,\,
-\xi
\right)~.}
\end{eqnarray}
We have determined the vacuum expectation values $\langled~\rangled$
\begin{eqnarray}
\langled\CF_{\underline{i}\underline{i}}\rangled=
-2\left(\frac{e}{m}\pm i\frac{\xi}{m}\right)
\label{<F>}
\end{eqnarray}
and thus
\begin{eqnarray}
\langled g_{\underline{i}\underline{i}}\rangled=\mp 2\frac{\xi}{m}~,~~~~
\langled\boldsymbol{D}^{\underline{i}}\rangled=\frac{m}{\sqrt{N}}(
0,-i,\pm 1)~.
\end{eqnarray}
The positivity of the metric implies 
$\mp\frac{\xi}{m}>0$,
so that
on the vacua, we have 
\begin{eqnarray}
\langled\CV\rangled =\mp 2m\xi=2|m\xi|.
\end{eqnarray}

\subsection{Gauge symmetry breaking}

Let $\langled A\rangled$
be 
\begin{eqnarray}
\langled A\rangled=
{\rm diag}(\stackrel{N_1}{\overbrace{\lambda^{(1)},\cdots,\lambda^{(1)}}},
\stackrel{N_2}{\overbrace{\lambda^{(2)},\cdots,\lambda^{(2)}}},\cdots)~~,
~~~~~\sum_iN_i=N
\end{eqnarray}
where $\lambda^{(k)}$ are complex roots of (\ref{<F>}),
then
$U(N)$ is broken to $\prod_{i} U(N_i$),
because
\begin{eqnarray}
[E^\pm_{\underline{j}\underline{k}},\langle A\rangle]
=\mp i(\lambda^{\underline{j}}-\lambda^{\underline{k}})E^\mp_{\underline{j}\underline{k}}~.
\end{eqnarray}
Unbroken $\displaystyle\Pi_i U(N_i$) is generated by
$t_\alpha\in \{t_a|[t_a,\langled A\rangled]=0\}$,
while broken ones by $t_\mu\in \{t_a|[t_a,\langled A\rangled]\neq 0\}$.
As will be seen in the next section, 
the mass spectrum is expressed in terms of unbroken $t_\alpha$ and broken $t_\mu$,
only. 
For later use, we note that
for a given $t_\mu$, there exists a unique $t_{\tilde \mu}$ 
such that $[t_\mu,\langle A\rangle]\sim t_{\tilde\mu}$,
which implies that
$f^{\tilde\mu}_{\mu\underline{i}}\lambda^{\underline{i}}=
-f^{\mu}_{\tilde\mu\underline{i}}\lambda^{\underline{i}}
$~.

\subsection{Partial supersymmetry breaking}

The supersymmetry transformation of fermions reduces on the vacua to
\begin{eqnarray}
\langled\boldsymbol{\delta \lambda}_I{}^{\underline{i}}\rangled
=i\langled(\boldsymbol{\tau\cdot  D}^{\underline{i}})_I{}^J\rangled
\boldsymbol{\eta}_J
~~~~~~~\mbox{while}~~~~
\langled\boldsymbol{\delta \lambda}_I{}^{r}\rangled=0
\end{eqnarray}
because $\langled\boldsymbol{D}^{r}\rangled=0$.
Note that {$\langled \det (\boldsymbol{\tau\cdot D}^{\underline{i}})\rangled
=-\langled\boldsymbol{ D }^{\underline{i}}\cdot \boldsymbol{D}^{\underline{i}}\rangled=0$},
thus supersymmetry is partially broken on the vacua.
In fact
\begin{eqnarray}
\langled\frac{1}{\sqrt{2}}\boldsymbol{\delta}
(\lambda^{\underline{i}}\pm \psi^{\underline{i}})\rangled
=\pm im\sqrt{\frac{2}{N}}(\eta_1\mp\eta_2)~,
~~~~
\langled\frac{1}{\sqrt{2}}\boldsymbol{\delta}
(\lambda^{\underline{i}}\mp \psi^{\underline{i}})\rangled
=0~.
\end{eqnarray}
The former implies that 
{$\langled\frac{1}{\sqrt{2}}\boldsymbol{\delta}
(\lambda^{i}\pm \psi^{i}) \rangled=\pm 2im\delta^i_0(\eta_1\mp\eta_2)$}.
As will be seen soon, 
$\frac{1}{\sqrt{2}}
(\lambda^{0}\pm \psi^{0})$ is massless and thus is the Nambu-Goldstone (NG) fermion
for the partial breaking of $\CN=2$ supersymmetry to $\CN=1$. 
We note that partial supersymmetry breaking is possible only if
$\im \langled \boldsymbol{D^a}\rangled\neq 0$.

\section{Mass spectrum}

\subsection{fermion mass spectrum}
We find that the fermion mass term
reduces to
\begin{eqnarray}
\langled \CL_{\rm Yukawa}\rangled&=&
\frac{1}{2}
\boldsymbol{\lambda}^{\alpha I}
(M_{\alpha\alpha})_I{}^J
\boldsymbol{\lambda}^{\alpha}_J
+
\frac{1}{2}
\boldsymbol{\lambda}^{\mu I}
(M_{\mu\nu})_I{}^J
\boldsymbol{\lambda}^{\nu}_J
~~
+c.c.\\
M_{\alpha\alpha}&=&
\frac{\sqrt{N}}{2}\langled \CF_{0\alpha\alpha}\rangled
(\boldsymbol{\tau}\cdot \langled \boldsymbol{D}^{\alpha}\rangled)
=
\frac{m}{2}\langled\CF_{0\alpha\alpha} \rangled
\left(
  \begin{array}{cc}
    \pm 1   &-1    \\
    1   &\mp 1    \\
  \end{array}
\right)
~,\label{fermion:alpha}\\
M_{\mu\tilde\mu}&=&
-M_{\tilde\mu\mu}=
\frac{1}{\sqrt{2}}\langled g_{\mu\mu}\rangled
f^{\mu}_{\tilde\mu\underline{k}}
\lambda^{*\underline{k}}
\left(
  \begin{array}{cc}
   1    &0    \\
    0   &1    \\
  \end{array}
\right)~.
\end{eqnarray}
Because $\det M_{\alpha\alpha}=0$,
the fermions $\boldsymbol{\lambda}^\alpha_J$ contain
massless modes,
while all of the fermions $\boldsymbol{\lambda}^\mu_J$
are massive.
Taking the normalization of the kinetic terms into account, 
we find 
fermion masses
on the vacua
\begin{eqnarray}
  \begin{array}{c|c|c|c}
 \mbox{field}      &\mbox{mass}    & \mbox{label}    &\mbox{\# of polarization states } \\
 \hline
\frac{1}{\sqrt{2}}(\lambda^\alpha\pm\psi^\alpha)       &0    &\mbox{A}    &2 d_u    \\
\frac{1}{\sqrt{2}}(\lambda^\alpha\mp\psi^\alpha)       
&| m\langled g^{\alpha\alpha}\CF_{0\alpha\alpha}\rangled |    &\mbox{B}    &2 d_u    \\
\boldsymbol{\lambda}^\mu_I       &\frac{1}{\sqrt{2}}|f^{\tilde \mu}_{\mu \underline{i}}\lambda^{*\underline{i}}|    &\mbox{C}    &4(N^2-d_u)    \\
  \end{array}
\label{fermion masses}
\end{eqnarray}
where {$d_u\equiv \dim \prod_iU(N_i$)}.
We obtain the 
NG
fermion $\frac{1}{\sqrt{2}}(\lambda^0\pm\psi^0)$
associated with the overall $U(1)$ part.

\subsection{boson mass spectrum}
Gauge boson mass term emerges from the kinetic term
\begin{eqnarray}
-\langled\CL_{\rm kin} \rangled
=\frac{1}{4}\langled g_{aa'} \rangled f^a_{bc}v_m^b\lambda^c
f^{a'}_{b'c'}v_m^{b'}\lambda^{*c'}
=\frac{1}{4}|f^{\tilde \mu}_{\mu \underline{i}}\lambda^{\underline{i}}|^2v_m^\mu v^{m\mu}
~,
\end{eqnarray}
which implies that $v_m^\mu$ are massive while
$v_m^\alpha$ massless.
The scalar mass term is extracted
by substituting
$A^a=\langled A^a\rangled +\delta A^a$
into $\CV$
\begin{eqnarray}
&&
\langled\partial_a\partial_{b^*}\CV\rangled \delta A^a\delta A^{*b}
+\frac{1}{2}\langled\partial_a\partial_{b}\CV\rangled \delta A^a\delta A^{b}
+\frac{1}{2}\langled\partial_{a^*}\partial_{b^*}\CV\rangled \delta  A^{*a}\delta A^{*b}
\equiv
\frac{1}{2}\overrightarrow{\delta A^a}^\dag
M_{ab}
\overrightarrow{\delta A^b}
\nonumber\\&&
\end{eqnarray}
where 
{$\overrightarrow{\delta A^a}\equiv
\binom{\delta A^a}{\delta A^{*a}}
$} and
\begin{eqnarray}
M_{\alpha\alpha}&=&m^2\langled g^{\alpha\alpha} |
\CF_{0\alpha\alpha}|^2\rangled
\left(\!\!
  \begin{array}{cc}
    1   &0    \\
      0 & 1   \\
  \end{array}
\!\!\right)~,
\\
M_{\mu\mu}&=&
\frac{1}{2}\langled g_{\tilde\mu\tilde\mu} \rangled 
\left(\!\!
  \begin{array}{cc}
|f^{\tilde \mu}_{\mu \underline{i}}\lambda^{\underline{i}}|^2       
 &-(f^{\tilde \mu}_{\mu \underline{i}}\lambda^{\underline{i}})^2    \\
-(f^{\tilde \mu}_{\mu \underline{i}}\lambda^{*\underline{i}})^2       
 &|f^{\tilde \mu}_{\mu \underline{i}}\lambda^{\underline{i}}|^2    \\
  \end{array}
\!\!\right)
=
\frac{1}{2}\langled g_{\tilde\mu\tilde\mu} \rangled 
T^{-1}
\left(\!\!
  \begin{array}{cc}
0     
 &0    \\
0      
 &2|f^{\tilde \mu}_{\mu \underline{i}}\lambda^{\underline{i}}|^2    \\
  \end{array}
\!\!\!\right)T~.~~~~
\end{eqnarray}
The massless mode $
(T
\overrightarrow{\delta A^\mu})_1$
is absorbed into $v_m^\mu$ as the longitudinal mode to form massive vector fields.
The resulting boson mass spectrum 
is summarized as
\begin{eqnarray}
  \begin{array}{c|c|c|c}
 \mbox{field}      &\mbox{mass}    & \mbox{label}    &\mbox{\# of polarization states } \\
 \hline
v_m^\alpha     &0    &\mbox{A}    &2 d_u    \\
A^\alpha    
&| m\langled g^{\alpha\alpha}\CF_{0\alpha\alpha}\rangled |    &\mbox{B}    &2 d_u    \\
v_m^\mu       &\frac{1}{\sqrt{2}}|f^{\tilde \mu}_{\mu \underline{i}}\lambda^{\underline{i}}|    &\mbox{C}    &3(N^2-d_u)    \\
(T \overrightarrow{A^\mu})_2 
&\frac{1}{\sqrt{2}}|f^{\tilde \mu}_{\mu \underline{i}}\lambda^{\underline{i}}|    
&\mbox{C}    &N^2-d_u    \\
  \end{array}
  \label{boson masses}
\end{eqnarray}
\bigskip

Due to the $\CN=1$ supersymmetry on the vacua,
the modes
in (\ref{fermion masses}) and (\ref{boson masses})
 form $\CN=1$ multiplets as follows.
First, fields labelled by A,
$(\frac{1}{\sqrt{2}}(\lambda^\alpha\pm\psi^\alpha), v_m^\alpha)$,
form
massless $\CN=1$ vector multiples of spin($\frac{1}{2},1$).
Secondly,
those labelled by B,
$(A^\alpha, \frac{1}{\sqrt{2}}(\lambda^\alpha\mp\psi^\alpha))$,
form massive $\CN=1$ chiral multiplets of spin($0,\frac{1}{2}$).
Finally those labelled by C,
$((T\overrightarrow{A^\mu})_2, \boldsymbol{\lambda}^\mu_I, v_m^\mu)$,
form two massive $\CN=1$ vector multiplets of spin($0,\frac{1}{2},1$).
The masses for these multiplets are depicted in Figure 1. 
\begin{figure}[h]
   \begin{center}
      \psfig{file=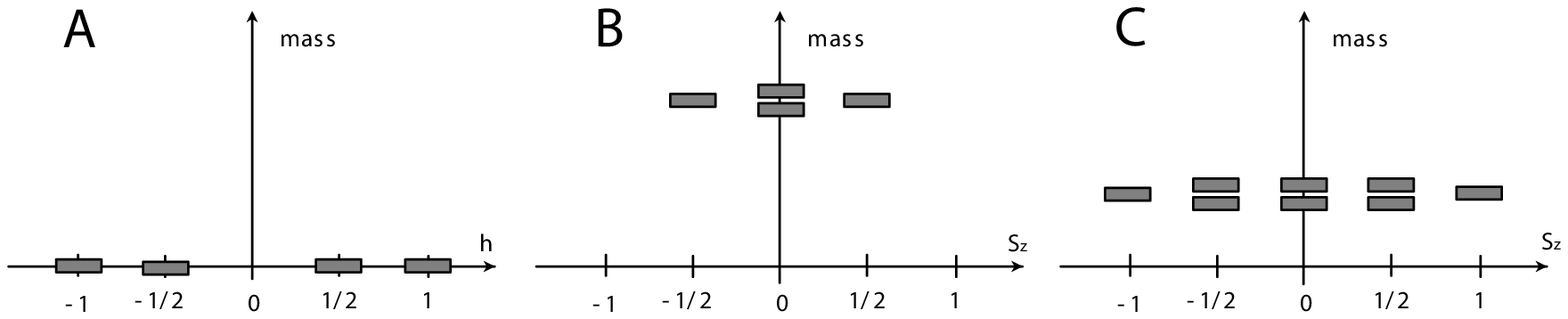,height=28mm}
   \end{center}
   \caption{$\CN=1$ supermultiplets}
\end{figure}

\section{Supercurrents and ``central charge''}
$U(1)_R$ transformation is given by
\begin{eqnarray}
R:~\Phi(x,\theta,\bar\theta)
\to e^{i\alpha}\Phi(x,e^{-i\frac{\alpha}{2}}\theta,\bar\theta),~~~
\CW_\alpha(x,\theta,\bar\theta)
\to \CW_\alpha(x,e^{-i\frac{\alpha}{2}}\theta,\bar\theta).
\end{eqnarray}
If $\CF$ transforms as weight two,~$R: \CF\to e^{2i\alpha}\CF$,
then our action is invariant under $R$ and 
the associated $U(1)_R$ current $J^m$ is
\begin{eqnarray}{
\theta \sigma_m \bar\theta~ J^m\equiv
\theta J \bar\theta=
(\tau_2)_{ab}\left(
\bar\theta\boldsymbol{\bar\lambda}^{Ia}~
 \theta\boldsymbol{\lambda}_I{}^{b}
+iA^{*a}\theta
\overleftrightarrow{\CD}_m\sigma^m\bar\theta A^b
\right)
\equiv
(\tau_2)_{ab}\left(
\bar\theta j^{ab}
 \theta
\right)~.}~~~
\end{eqnarray}
Acting a supersymmetry transformation
on this we obtain
$\CN=2$ supercurrents~\cite{Ferrara:1974pz}\cite{Itoyama:1996hh}
\begin{eqnarray}{
\boldsymbol{\eta}_J\boldsymbol{S}^{(J)m}+
\boldsymbol{\bar \eta}_J\boldsymbol{\bar S}^{(J)m}
=-\frac{1}{2}(\tau_2)_{ab}~\tr (\bar\sigma^m\, \boldsymbol{\delta}j^{ab})~}
\end{eqnarray}
where $\tr$ is for spinor indices.

Though the $R$-current is not conserved for
general $\CF$, 
we can construct a conserved $\CN=2$ supercurrent as
a broken $\CN=2$ supermultiplet of currents.
We write 
{$\CF=\sum_{n} h_nC^{(n)}(A^a)$} where 
$C^{(n)}(A^a)$ are $n$-th order $U(N)$ invariant polynomials 
in $A^a$
and $h_n$ are their coefficients.
First, we assign weight $(2-n)$ to $h_n$.
Then the weight of $\CF$ can be regarded as two.
The local $U(1)_R$ variation of $\CL$
implies
\begin{eqnarray}
{\partial_m 
\left
( -
\frac{1}{2}
\tr \bar{\sigma}^m J
\right
)
=i \left
( \sum_{n} (n-2) \frac{\partial}{\partial h_n} 
\right
) \mathcal{L} \equiv \Delta_h \mathcal{L}~.}
\end{eqnarray}
Acting the supersymmetry transformation on it,
and noting that $\boldsymbol{\delta}\CL=\partial_m X^m$ with
some $X^m$ and that $\Delta_h\partial_mX^m=\partial_m \Delta_h X^m$,
we obtain  a general construction of the conserved 
$\mathcal{N}=2$ supercurrents of our model;
\begin{eqnarray}
{\boldsymbol{\eta}_J \boldsymbol{\mathcal{S}}^{(J)m}+
\bar{\boldsymbol{\eta}}^J \bar{\boldsymbol{\mathcal{S}}}_{(J)}^{\ m} \equiv 
-{\frac{1}{2}} \tr (\bar{\sigma}^m \boldsymbol{\delta}J)-\Delta_h X^m . }
\end{eqnarray}
The term $\Delta_h X^m$ is not universal but
depends on the concrete form of $\CF$.
It should be difficult to find the universal coupling to supergravity.

Further action of the supersymmetry transformation
generates
\begin{eqnarray}
{\theta\boldsymbol{\delta \delta} J \bar\theta =
8m\xi ~\bar\theta \boldsymbol{\bar\eta}\,
\boldsymbol{\tau_1}\,\boldsymbol{\eta}\theta
~+~\cdots ,}
\end{eqnarray}
from which we can read off the constant matrix
$C_I{}^J$ in (\ref{supercurrent algebra})
as $C_I{}^{J}=+4m\xi (\boldsymbol{\tau}_1)_I{}^{J}$.

\section{$\CN=2$ $U(N)$ gauge model coupled with $\CN=2$ hypermultiplets}
In this section we provide a manifestly $\CN=2$ supersymmetric
formulation of the $\CN=2$ $U(N)$ gauge model
coupled with $\CN=2$ hypermultiplets\cite{FIS4}.
For this we work in harmonic superspace\cite{HS} 
$\bR^{4|8}\times S^2$
parametrized by
\begin{eqnarray}
(x^m_A,\theta^\pm,\bar\theta^\pm,u_I^\pm)
=
(x^m-2i\theta^I\sigma^m\bar\theta^Ju_{(I}^+u_{J)}^-,
\theta^Iu_I^\pm,
\bar\theta^Iu_I^\pm,
u_I^\pm)~
\end{eqnarray}
in the analytic basis.
$u_I^\pm$ are  harmonic variables parametrizing S$^2=SU(2)/U(1)$
\begin{eqnarray}
(u_I^+,u_I^-)\in SU(2)~,~~
u^{+I}u_I^-=1~,~~
\overline{u^{+I}}=u_I^-~.
\end{eqnarray}

We introduce an $\CN=2$ vector multiplet
$V^{++}(x^m, \theta^+,\bar\theta^+)=V^{++a}t_a$
transforming as adjoint under $U(N)$.
$V^{++}$ is composed of
a complex scalar $A$,
a vector $v_m$,
an $SU(2)$ doublet Weyl spinor $\lambda^{i}_\alpha$
and an auxiliary field $D^{IJ}$.
$D^{I}{}_J=\varepsilon_{JK}D^{IK}= i D^A(\tau_A)^I{}_J$
is an $SU(2)$ matrix
and $D^A$ is a real three-vector $\overline{D^A}=D^A$.
By using the field strength $W$ of $V^{++}$
the action for $V^{++}$ is
constructed as
\begin{eqnarray}
S_V&=&-\frac{i}{4}\int d^4x\left[
(D)^4\CF(W)-(\bar D)^4\bar\CF(\bar W)
\right]~,
\end{eqnarray}
where $(D)^4=\frac{1}{16}(D^+)^2(D^-)^2$ and
$D^\pm$ are the spinor harmonic derivatives\cite{HS}. 
$A$'s parametrize the special K\"ahler geometry
with the K\"ahler metric $g_{ab}=\im(\CF_{ab}|)$
where
$\CF_{ab\cdots}|$ means $\CF_{ab\cdots}$ evaluated at $\theta^\pm=\bar\theta^\pm=0$.
The metric admits $U(N)$ isometry generated by Killing vectors
with the Killing potential 
$\CP^a=-if^a_{bc}\bar A^bA^c$.

The $\CN=2$ hypermultiplet $q^+$ 
is composed of
an $SU(2)$ doublet complex scalar $f^I$,
a pair of $SU(2)$ singlet spinors
and
infinitely many auxiliary fields.
We introduce two sets of $\CN=2$ hypermultiplets,
$N_f$ hypermultiplets $q^{+u}$ ($u=1,\cdots,N$)
and $N_a$ hypermultiplets $q^{+a}$ ($a=0,1,\cdots,N^2-1$)
which transform
as fundamental representation
and adjoint representation under $U(N)$, respectively.
We suppress flavor indices below.
The $U(N)$ gauged action is given by
($\omega$-hypermultiplets are also included in ref.\citen{FIS4})
\begin{eqnarray}
S_q^{\rm gauged}&=&
-\int dud\zeta^{(-4)}\left[
\tilde q^+_u\CD^{++}q^{+u}
+\tilde q^+_a\CD^{++}q^{+a}
\right]~
\end{eqnarray}
where the tilde denotes the analyticity preserving conjugation\cite{HS}.
The covariant derivative is defined as
$
\CD^{++}q^{+\mu}=
D^{++}q^{+\mu}+iV^{++a}(T_a)^\mu{}_\nu q^{+\nu}
$
where $D^{++}$ is the harmonic derivative\cite{HS}
and
\newcommand\ad{{\rm ad}}
\begin{eqnarray}
(T_a)^\mu{}_\nu=
\left\{\!\!\!
  \begin{array}{ll}
(t_a)^u{}_v       &\mbox{for fundamental} ~q^+     \\
\ad(t_a)^b{}_c=if^b_{ac}       &\mbox{for adjoint} ~q^+    \\
  \end{array}
\right.
.
\end{eqnarray}
The $U(N)$ isometry gauged above
is generated by  Killing vectors with
Killing potentials
\begin{eqnarray}
\left\{
  \begin{array}{l}
\hat \CQ^{IJ}_a=\CQ^{IJ}_a|_{T_a=t_a}       \\
\check  \CQ^{IJ}_a=\CQ^{IJ}_a|_{T_a=\ad(t_a)}       \\
  \end{array}
\right.
~~\mbox{where}~~~
\CQ^{IJ}_a=
i
\bar f^{(I}_\mu(T_a)^\mu{}_\nu f^{J)\nu}
~.
\end{eqnarray}

Next we introduce
the electric and magnetic FI terms.
The electric FI term is given by
\begin{eqnarray}
S_e&=&
\int dud\zeta^{(-4)}\tr(\Xi^{++}V^{++})+c.c.
=\int d^4x \xi^{IJ}D^0_{IJ}+c.c.
\end{eqnarray}
where $\Xi^{++}=\xi^{IJ}u_I^+u_J^+$ is the electric FI parameter.
The effect of this term is to shift the dual auxiliary field $D_D^{aIJ}$
in $W_D^a\equiv \CF_a$
by an imaginary constant,
$D_D^{aIJ}\to D_D^{aIJ}+8i\xi_D^{IJ}\delta_0^a$.
We introduce the magnetic FI term
so as to shift the auxiliary field $D^{aIJ}$ in $W^a$
by an imaginary constant,
$D^{aIJ}\to \boldsymbol{D}^{aIJ}=D^{aIJ}+4i\xi^{IJ}\delta_0^a$.
By this, the $\CN=2$ supersymmetry transformation law
$\delta_\eta\lambda^{aI}=(D^a)^I{}_J\eta^J+\cdots$
changes to 
$\delta_\eta\lambda^{aI}=(\boldsymbol{D}^a)^I{}_J\eta^J+\cdots$,
under which
the total action
\begin{eqnarray}
S=S_V+S^{\rm gauged}_q+S_e+S_m
\end{eqnarray}
is invariant.
It is straightforward to see that
the magnetic FI term of the form
\begin{eqnarray}
S_m&=&\int d^4x
\left[ (D)^4
\xi_D^{IJ}\theta_I\theta_J
\left(
\CF_0+2i\CF_{00}\xi_D^{KL}\theta_K\theta_L
\right)
+2i\hat \CQ^{IJ}_0\xi_{DIJ}
\right]
+c.c.
\label{magFI}
\end{eqnarray}
causes the imaginary constant shift of the auxiliary field
\begin{eqnarray}
S_V+S^{\rm gauged}_q +S_m
=
\left.\left(
S_V+S^{\rm gauged}_q
\right)\right|_{D\to\boldsymbol{D}}~
\end{eqnarray}
where
$|_{D\to\boldsymbol{D}}$ means the replacement
$D^{aIJ}\to\boldsymbol{D}^{aIJ}$ ($D^{aIJ}\to\bar{\boldsymbol{D}}^{aIJ}$).
We find that
in the presence of hypermultiplets in the fundamental representation
of the gauge group $U(N)$,
the magnetic FI term 
develops an additional term
which overcomes the difficulty in coupling
fundamental hypermultiplets with the APT model.
The adjoint scalars do not appear in (\ref{magFI})
because $\ad(t_0)=0$.

It is straightforward to eliminate infinitely many auxiliary fields 
in $q^+$
and the auxiliary field $D$ in $V^{++}$
and obtain the scalar potential
\begin{eqnarray}
\CV&=&
\frac{1}{4}g_{ab}\boldsymbol{D}^{aIJ}|\bar{\boldsymbol{D}}^{b}_{IJ}|
+g_{ab}\CP^a\CP^b
+2i(\xi^{IJ}+\bar\xi^{IJ})(\xi_{DIJ}-\bar\xi_{DIJ})
\nonumber\\&&
+\overline{f^I}_u(\bar AA+A\bar A)^u{}_vf^{Iv}
+\overline{f^I}_a(\bar AA+A\bar A)^a{}_bf^{Ib}
\end{eqnarray}
where
\begin{eqnarray}
\boldsymbol{D}^{aIJ}|=
-2g^{ab}\left[
(\xi^{IJ}+\bar\xi^{IJ})\delta_b^0
+(\xi_D^{IJ}+\bar\xi_D^{IJ})\bar\CF_{0b}|
+\hat \CQ^{IJ}_b+\check \CQ^{IJ}_b
\right]~.
\nonumber
\end{eqnarray}
The vacua are determined by $\CV$
and exhibit various phases.
On the Coulomb phase 
$\langled A^{\underline{i}}\rangled\neq 0$,
$\langled A^{r}\rangled=
\langled f^I_u\rangled=
\langled f^I_r\rangled=0$
and thus $\langled \hat\CQ_b^{IJ}\rangled=
\langled\check\CQ_b^{IJ}\rangled=0$.
In this way we have arrived at 
the vacuum condition
for $\CN=2$ $U(N)$ gauge model without hypermultiplets,
$
\langled\partial_{A^a}\CV\rangled=
\frac{i}{4}\langled\CF_{abc}|\boldsymbol{D}^{bA}\boldsymbol{D}^{cA}\rangled
=0
$.
Let us examine the case with $\CF=\sum_n\frac{c_n}{n!}\tr W^n$
for concreteness. Then it further reduces to
\begin{eqnarray}
\sum_A\langled\boldsymbol{D}^{\underline{i}A}\boldsymbol{D}^{\underline{i}A}\rangled=0~,~~~
\underline{i}=\underline{1},\cdots,\underline{N}~.
\label{vacuum condition reduced}
\end{eqnarray}
It is easy to show that by fixing $SU(2)$ appropriately
$\langled\CF_{\underline{i}\underline{i}}\rangled
=-2(\frac{e}{m}\pm i\frac{\xi}{m})$ in (\ref{<F>})  
can be reproduced.
On the vacua the supersymmetry transformations
of fermions are found to be trivial
except for
\begin{eqnarray}
\langled\delta\lambda^{\underline{i}I}\rangled=
i\langled\boldsymbol{D}^{\underline{i}A}\rangled(\tau_A)^I{}_J\eta^J~.
\end{eqnarray}
On the other hand the mass term of $\lambda^{\underline{i}I}$
is
\begin{eqnarray}
-\frac{i}{4}\langled\CF_{\underline{i}\underline{i}\underline{i}}
\boldsymbol{D}^{\underline{i}A}\rangled
\lambda^{\underline{i}I}
(\tau_2\tau_A)_{IJ}
\lambda^{\underline{i}J}~.
\end{eqnarray}
Because (\ref{vacuum condition reduced}) implies that 
$\det \boldsymbol{D}^{\underline{i}}{}^I{}_J=0$,
a half of the fermions $\lambda^{\underline{i}I}$, $I=1,2$,
say $U^1{}_J\lambda^{\underline{i}J}$ with a constant matrix $U^I{}_J$, 
is massless but has a nontrivial supersymmetry transformation.
In the ordinary basis spanned by matrices $t_a$, this means
that $U^1{}_J\lambda^{0J}$ is the NG fermion for
partial supersymmetry breaking.

\section*{Acknowledgements}
This work is supported in part by the Grant-in-Aid for Scientific
Research
(No.16540262, No.17540262 and No.17540091) 
from the Ministry of Education,
Science and Culture, Japan.
Support from the 21 century COE program
``Constitution of wide-angle mathematical basis focused on knots"
is gratefully appreciated.

\end{document}